\documentclass[11pt]{article}
\usepackage{multicol}

\usepackage{color,cite,amsbsy,amssymb,amsmath}
\usepackage{graphics,graphicx,ulem}

\newcommand{\rem}[1]{}
\newcommand{\bom}{\mbox{\boldmath$\omega$}}
\newcommand{\bu}{\mbox{\boldmath$u$}}
\newcommand{\bx}{\mbox{\boldmath$x$}}
\newcommand{\bdf}{\mbox{\boldmath$f$}}

\newcommand{\ID}{\int_{\mathcal{V_{D}}}}
\newcommand{\bel}{\begin{equation}\label}
\newcommand{\ee}{\end{equation}}
\newcommand{\beq}{\begin{eqnarray}\label} 
\newcommand{\eeq}{\end{eqnarray}} 
\newcommand{\bc}{\begin{center}} 
\newcommand{\ec}{\end{center}} 
\newcommand{\ben}{\begin{enumerate}}
\newcommand{\een}{\end{enumerate}}
\newcommand{\bit}{\begin{itemize}}
\newcommand{\eit}{\end{itemize}}
\newcommand\shalf{\ensuremath{{\scriptstyle\frac{1}{2}}}}

\newcommand\quart{\ensuremath{{\scriptstyle\frac{1}{4}}}}

\newtheorem{theorem}{Theorem}

\begin{document}

\bc
\textbf{\large\color{blue}Intermittency, cascades and thin sets in three-dimensional\\
Navier-Stokes turbulence}\\
\bigskip
\textbf{John D. Gibbon}\\
\bigskip
\textbf{Department of Mathematics,\\Imperial College London,\\
London SW7 2AZ, UK}
\ec

\begin{abstract}
Visual manifestations of intermittency in computations of three dimensional Navier-Stokes fluid turbulence appear as the low-dimensional or `thin' filamentary sets on which vorticity \& strain accumulate as energy cascades down to small scales. In order to study this phenomenon, the first task of this paper is to investigate how weak solutions of the Navier-Stokes equations can be associated with a cascade \&, as a consequence, with an infinite sequence of inverse length scales. It turns out that this sequence converges to a finite limit. The second task is to show how these results scale with integer dimension $D=1,\,2,\,3$ \&, in the light of the occurrence of thin sets, to discuss the mechanism of how the fluid might find the smoothest, most dissipative class of solutions rather than the most singular.
\end{abstract}




\begin{multicols}{2}

\section{\large Introduction}


\begin{figure*}[ht]
\bc
\includegraphics[scale=0.28]{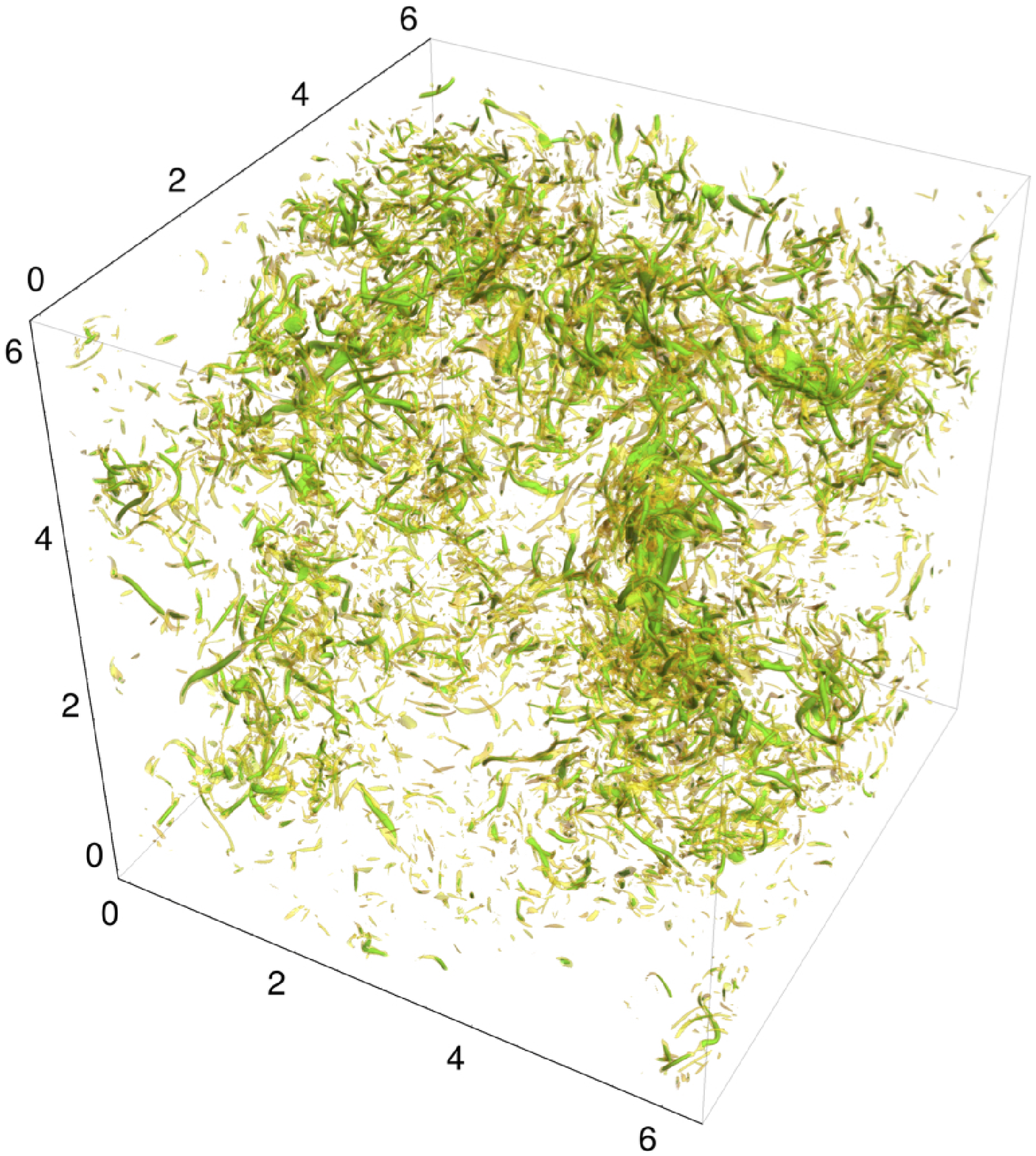}
\hspace{20mm}
\includegraphics[scale=0.28]{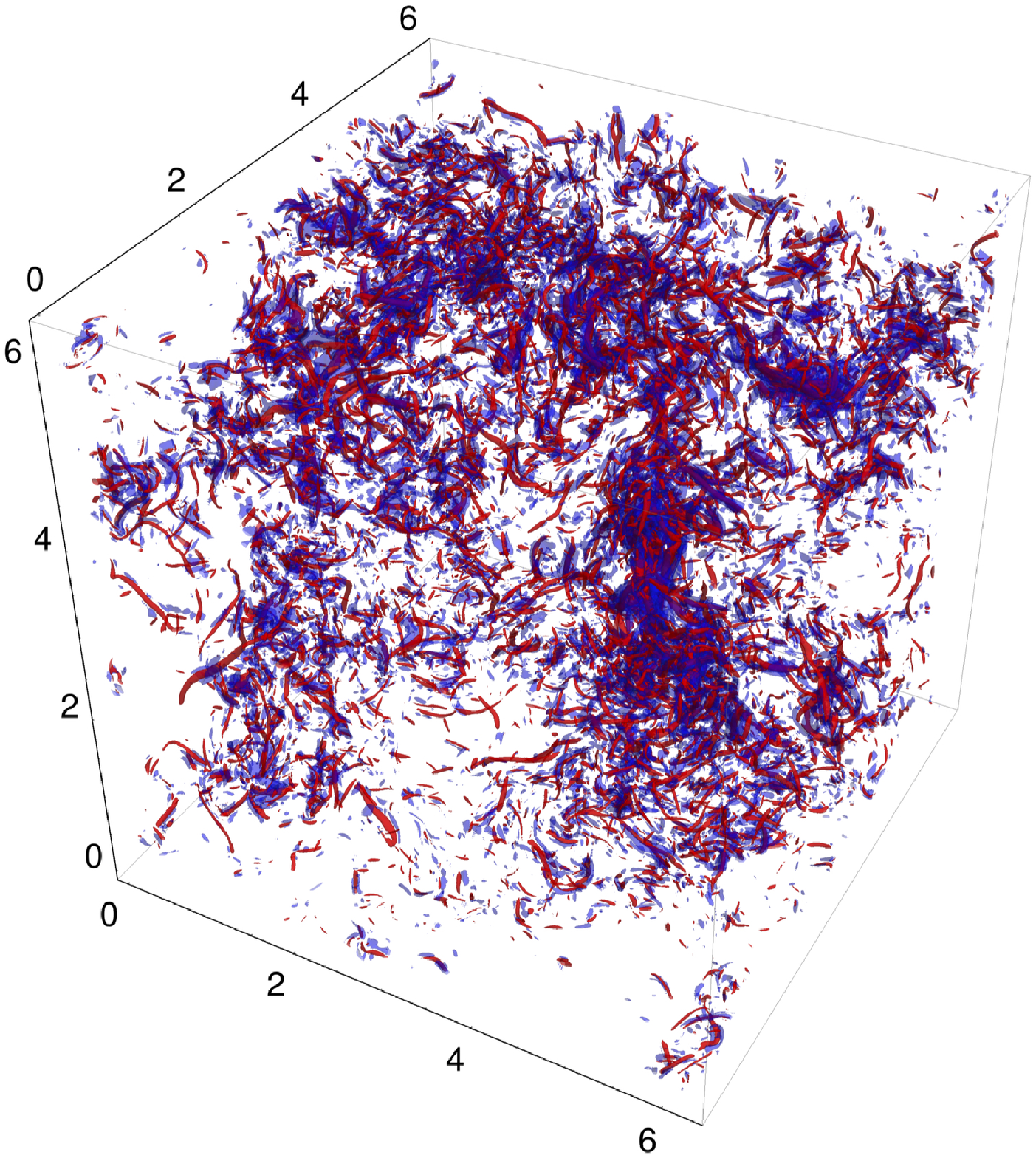}
\ec
\caption{The left-hand figure is a snapshot of the energy dissipation field $\varepsilon=2\nu S_{i,j}S_{j,i}$ of a forced $512^{3}$ Navier-Stokes flow at $Re_{\lambda} =196$ which is colour-coded such that yellow is 4 times the mean \& blue denotes 6 times the mean.  The right-hand figure shows the field $Q = \shalf\left(|\bom|^{2}-|S|^{2}\right)$\,: the colours correspond to  $-2Q_{rms}$ (blue) \& $5Q_{rms}$ (red). Plots courtesy of J. R. Picardo \& S. S. Ray.}
\end{figure*}
The most striking visual manifestation of intermittency in three-dimensional incompressible fluid turbulence is the accumulation of vorticity \& strain on `thin' or low-dimensional sets. When displayed graphically as iso-surfaces in a cube, these sets typically appear as spaghetti-like entangled tubular filaments\,: see Fig. 1 where snapshots of the energy dissipation field $\varepsilon$, \& the $Q$-field (defined in the caption) of a forced Navier-Stokes flow are displayed. Although differing in the fine detail from case to case, initial three-dimensional vortical structures tend to flatten at intermediate times into quasi-two-dimensional pancakes which subsequently roll up into quasi-one-dimensional tubes, with further iterations of flattening \& filamentation resulting in ever finer striations \cite{DCB1991,TK1993,JWSR1993,VM1994,MKO1994,DYS2008,IGK2009,EM2010, HIK2013,HIWK2014,EIGSH2017}\,: see the recent paper by Elsingha, Ishihara \& Hunt \cite{EIH2020}. Historically, Batchelor and Townsend \cite{BT1949} were the first to suggest that vorticity and strain are not distributed in a Gaussian fashion across a domain but accumulate on local, intense sets which they identified with intermittency in the energy dissipation \cite{KC1971,SM1991,UF1995,BB2009,MKV2020}. In the literature these filamentary structures are loosely referred to as `fractal' because of the roughness of the detail of their evolving fine-scale structure\,: see Fig. 1 for an illustration. In the last generation, various cascade models, such as the beta, bi-fractal and multi-fractal models, explicitly talk about accumulation on sets of non-integer dimension $D$ \cite{UF1995}. Studies in \textit{Fourier decimation} have pursued the idea of intermittency more precisely by projecting a three dimensional Navier-Stokes velocity field onto a chosen subset of Fourier modes by employing a generalized Galerkin projector \cite{FPR2012,Ray2015}. Intermittency properties have then been investigated by tuning both the restricted subset and the Reynolds number. Recent work has culminated in making this restricted set fractal \cite{LBMTB2015,BBFR2016,LMB2016,BBBLR2016}\,: in effect, the number of degrees of freedom are limited to a sphere of radius $k$ growing as $k^{D}$ (for non-integer $D$) embedded in the  three-dimensional space. Ref. \cite{BBBLR2016} contains an excellent set of references. However, from the point of view of rigorous Navier-Stokes analysis, the phenomenon is by no means understood, mainly because technical tools exist to pursue analysis only on fixed domains of integer dimension but not on time-evolving fractal sets. In this context, many questions remain outstanding. Is there a universal value of $D$ or are there many disjoint sets of differing dimension? How does this fit in with the idea of a cascade to small scales and the regularity of solutions of the Navier-Stokes equations at these scales? For instance, Biferale and Titi \cite{BT2013} have shown that a helically decimated version of the $3D$ Navier-Stokes equations leads to global regularity. A global theory that answers all these questions still remains elusive\,: this paper aims to build on what is known rigorously for three dimensional Navier-Stokes equations to discuss how this fractal set might occur.

\section{\large Cascades, scaling \& weak solution estimates in three dimensions}

A cascade is a sequential process that involves vorticity \& strain being driven down to ever smaller length scales in the flow  
\& has long been closely associated with intermittency \cite{UF1995,Ray2018,BMT2012,BMT2013}. For sufficiently long times a cascade to smaller scales should show up in estimates of both spatially \& temporally averaged gradients of a divergence-free velocity field $\bu(\bx,\,t)$ that evolve according to the Navier-Stokes equations 
\bel{NS1}
\left(\partial_{t}+\bu\cdot\nabla\right)\bu + \nabla p =  \nu\Delta\bu + \bdf(\bx)\,.
\ee
The domain $\mathcal{V}=[0,\,L]^{3}_{per}$ is chosen to be three-dimensional \& periodic. $\nu$ is the viscosity \& $\bdf$ is an $L^{2}$-bounded forcing. Here we show that Leray's weak solutions of the Navier-Stokes equations \cite{Leray1934} can be interpreted in terms of a cascade. 
\par\smallskip
We define a doubly-labelled set of norms in dimensionless form
\bel{NS2a}
F_{n,m} = \nu^{-1} L^{1/\alpha_{n,m}} \|\nabla^{n}\bu\|_{2m}\,,
\ee
where $\alpha_{n,m}$ is defined by
\bel{alphadef}
\alpha_{n,m} = \frac{2m}{2m(n+1)-3}\,.
\ee
The norm notation $\|\cdot\|_{2m}$ in (\ref{NS2a}) is defined by
\bel{NS2b}
\|\nabla^{n}\bu\|_{2m} = \left(\int_{\mathcal{V}}|\nabla^{n}\bu|^{2m}dV\right)^{1/2m}\,.
\ee
Higher values of $n$ allow the detection of smaller scales, while higher values of $m$ account for stronger deviations from the mean, with $m=\infty$ representing the maximum norm. 
\par\smallskip
The Navier-Stokes equations are well-known to possess the scale-invariance property 
\bel{uinv}
\bu(\bx,\,t) \to \lambda^{-1}\bu\left(\bx/\lambda,\,t/\lambda^{2}\right)\,,
\ee
for any value of the dimensionless parameter $\lambda$. Under this scaling the $F_{n,m}$ in (\ref{NS2a}) are invariant in $\lambda$ \& are thus invariant at every length \& time scale in the flow. This makes them invaluable as a tool for investigating a cascade of energy through the system. This is further illustrated by the fact that there exists a bounded, weighted, double hierarchy of their time averages 
\bel{NS3}
\left<F_{n,m}^{\,\alpha_{n,m}}\right>_{T} \leq c_{n,m}Re^{3}\left\}
\begin{array}{ll}
n \geq 1~&~1 \leq m \leq \infty\,,\\
n =0~ &~3 < m \leq \infty\,,
\end{array}\right.
\ee
as demonstrated in \cite{JDG2018}. The angled brackets $\left<\cdot\right>_{T}$ are defined by 
\bel{timav}
\left<\cdot\right>_{T} = T^{-1}\int_{0}^{T}\cdot\,dt\,,
\ee
\& the Reynolds number $Re$ by
\bel{Reydef}
Re = LU/\nu\quad\mbox{with}\quad U^{2} = L^{-3}\left<\|\bu\|_{2}^{2}\right>_{T}\,.
\ee
The physical meaning of the set of inequalities in (\ref{NS3}) can be illustrated thus. Consider the time-averaged energy dissipation rate defined in the conventional manner as $\varepsilon_{av} = \nu L^{-3}\left<\|\nabla\bu\|_{2}^{2}\right>_{T}$. Then in the case $n=m=1$, (\ref{NS3}) becomes 
\bel{epsA}
\varepsilon_{av} \leq c_{1,1} \nu^{3}L^{-4}Re^{3}\,.
\ee
The upper bound is recognizable as the same result derived by Kolmogorov's theory \cite{UF1995} \&, as we shall see below, leads to the well-known $Re^{3/4}$ estimate for the inverse Kolmogorov length. The double hierarchy displayed in (\ref{NS3}) furnishes us with bounds which generalize (\ref{epsA}) to all derivatives \& in every $L^{2m}$-norm. It is valid for Leray's weak solutions \& encapsulates all the known weak solution results in Navier-Stokes analysis \cite{JDG2018}. These are distributional in nature \& are not unique \& thus the result in (\ref{NS3}) falls short of a full regularity proof\,; i.e. existence \textit{\&} uniqueness of solutions. It was shown in \cite{JDG2018} that to achieve this would require 
\bel{NS4}
\left<F_{n,m}^{2\alpha_{n,m}}\right>_{T} < \infty\,. 
\ee
While it remains an open problem, there is no evidence that any bounds with the factor of 2 in the exponent exist. Indeed it is possible that weak solutions are all that are available. What has been deduced is that  (\ref{NS2a}) \& (\ref{alphadef}) lead to a definition of a set of inverse length scales $\ell_{n,m}^{-1}$
\bel{ls1a}
\left(L\ell_{n,m}^{-1}\right)^{n+1} := F_{n,m}\,,
\ee
whose estimated time averages are \cite{JDG2018}
\bel{ls1b}
\left<L\ell_{n,m}^{-1}\right>_{T} \leq c_{n,m}Re^{\frac{3}{(n+1)\alpha_{n,m}}} + O\left(T^{-1}\right)\,.
\ee
The exponents of $Re$ in the two cases $n=m=1$ \& $n,\,m\to\infty$ are
\beq{ls1c}
\left.\frac{3}{(n+1)\alpha_{n,m}}\right|_{n,m=1} &=& 3/4\,,\\
\lim_{n,m\to\infty} \frac{3}{(n+1)\alpha_{n,m}} &=& 3\,.\label{ls1d}
\eeq
The first result in (\ref{ls1c}) is consistent with the inverse Kolmogorov length while the second result in (\ref{ls1d}) \textit{implies that there exists a finite limit to the cascade process.} However, when $Re$ is large it does so at a level below molecular scales where the Navier-Stokes equations are not valid. Nevertheless it validates Richardson's original assertion that viscosity eventually terminates the cascade process \cite{UF1995,LFR}. 


\section{\large Estimates \& scaling in $D$-dimensions}\label{Dint}

Inequalities (\ref{NS3}) \& (\ref{ls1b}) are true for weak solutions in a $D=3$ domain. For integer values of $D = 2$ or $D=3$ on a periodic domain $\mathcal{V}_{D}$ the definition of (\ref{NS2a}) can be generalized to\footnote{When $D=1$ the Navier-Stokes equations make no sense unless the pressure \& divergence-free terms are removed, in which case we have Burgers' equation. The results expressed in $D$-dimensions with $D=1$ are valid for this.}
\bel{FnmDdef}
F_{n,m,D} = \nu^{-1} L^{1/\alpha_{n,m,D}} \|\nabla^{n}\bu\|_{2m}\,,
\ee
where $\alpha_{n,m}$ in (\ref{alphadef}) \& (\ref{NS3}) is replaced by
\bel{alphad1}
\alpha_{n,m,D} = \frac{2m}{2m(n+1)-D}\,.
\ee
The $F_{n,m,D}$ in (\ref{FnmDdef}) possess the same invariance properties as $F_{n,m}$ in (\ref{NS2a}). The details of the proof of (\ref{NS3}) has been generalized for the integer $D$-dimensional case using the same methods \& results as in three dimensions \cite{JDG2018}, although the calculation is far from straightforward\,: see the Appendix.
\begin{theorem}\label{thm1}
For $D=2,\,3$ and for $n \geq 1$ \& $1 \leq m \leq \infty$, the equivalent of (\ref{NS3}) is 
\bel{FnmDbnd}
\left<F_{n,m,D}^{(4-D)\alpha_{n,m,D}}\right>_{T} \leq c_{n,m,D}\,Re^{3}\,.
\ee
For $D=1$ the same result holds for Burgers' equation.
\end{theorem}
More than 40 years ago Fournier \& Frisch \cite{FF1978} introduced the idea of turbulence in $D$ dimensions where $D$ is no longer an integer but is restricted to the range $D \geq 2$. They achieved this by analytically continuing the Taylor expansion in time of the energy spectrum $E_{k}(t)$, assuming Gaussian initial conditions. Since then the idea of a non-integer dimension has taken root in the many papers on the beta, bi-fractal \& multi-fractal models \cite{UF1995,BB2009}. Can the Navier-Stokes estimates in (\ref{FnmDbnd}) be performed on a domain of non-integer dimension? In a fully rigorous sense, the answer is in the negative. For instance, there are no proofs of the Divergence Theorem or the Sobolev inequalities on fractal domains. Thus we can only claim the validity of Theorem \ref{thm1} for integer values of $D$. What the result does do, however, is show how the exponent of $F_{n,m,D}$ scales with integer values $D$. The surprising but crucial factor of $4-D$ in the exponent multiplying $\alpha_{n,m,D}$ deserves some remarks\,:
\ben\itemsep 0mm
\item When $D=3$, the factor of $4-D$ is simply unity \& (\ref{FnmDbnd}) reduces to (\ref{NS3})\,;

\item When $n=m=1$ this factor cancels to make $(4-D)\alpha_{1,1,D}=2$ for every value of $D$, as it should. It also furnishes us with the correct bound on the averaged energy dissipation rate $\varepsilon_{av}$. 

\item When $D=2$ we achieve the $2\alpha_{n,m,2}$ bound required for full regularity, as in (\ref{NS4}). Thus the case $D=2$ is critical for regularity, as is well-known \cite{DG1995,FMRT,RRS}. 
\een
\par\smallskip
As in Fig. 1, computations in \cite{IGK2009,EM2010,HIK2013,HIWK2014,EIGSH2017,EIH2020} have shown that the process of flattening \& filamentation results in ever finer striations as the flow progresses. This would indicate that the set(s) on which vorticity or strain are concentrated has a non-integer \& decreasing dimension. While we have no rigorous methods for proving the validity of (\ref{FnmDbnd}) when $D$ takes non-integer values, it raises the intriguing possibility that this may nevertheless be true. Certainly it is clear that when $D$ decreases in (\ref{FnmDbnd}) then the exponent $(4-D)\alpha_{n,m,D}$ of $F_{n,m,D}$ increases, which is the direction of more, not less, regularity. \textit{This suggests that a flow may adjust itself to find the smoothest, most dissipative set, not the most singular, on which to operate.} This runs counter to the traditionally held theory of viscous turbulence in which singularities have been long-st\&ing c\&idates as the underlying cause of turbulent dynamics \cite{BHP2016,HKM2019,JPR2020}, even though they must be rare events \cite{CKN1982,DG1995,FMRT,RRS}. Adjustment to find the smoothest, most dissipative set could be a way of the flow re-organizing \& regularizing itself to avoid singularities.

\rem{
None of the well-established theories are able to explain why thin sets appear to be the natural spatial structures in Navier-Stokes turbulent three dimensional domains. \rem{although attempts have been to discuss this in terms of inherent geometric structures based around the sign of the Laplacian of the pressure \cite{HWM1988,Haller2005,RBRG,RBR}.} In short, the traditional \& contrasting interpretations in terms of cascades through the scales, resulting in either termination by viscous dissipation or by rapid singularity formation, do not explain the low-dimensional nature of the phenomenon. This paper is an attempt to address this problem. Firstly it shows how Leray's weak solutions of the Navier-Stokes equations \cite{Leray1934} can be interpreted in terms of a cascade, \& it then suggests a different accumulation mechanism, which travels in the opposite direction to the idea of singularity formation \cite{BHP2016,HKM2019}. }
\par\vspace{2mm}\noindent
\textbf{Acknowledgments\,:} I thank J. R. Picardo (IIT Mumbai) \& S. S. Ray (ICTS Bangalore) for the plots in Fig.1 from their Navier-Stokes data.
\end{multicols}

\section{\large Appendix\,: proof of Theorem 1}

The aim of Theorem \ref{thm1} is to roll together estimates for the Navier-Stokes equations that are already known individually in both the $D=2$ \& $D=3$ cases. In addition, Burgers' equation is included, which is appropriate for $D=1$ when the pressure term \& the incompressibility condition have been dropped. The main foundation of the proof of Theorem \ref{thm1} is the original result of  Foias, Guillop\'e \& Temam (FGT) in $3$-dimensions \cite{FGT1981}. Given that all three results are known separately, we are able to formally manipulate \& differentiate the $H_{n}$, defined below in (\ref{Hn1def}) below, on a periodic domain of integer dimension $D$.

\section{\small The FGT result in integer $D$ dimensions}\label{fgtsub}

We require the definition
\bel{Hn1def}
H_{n} = \ID |\nabla^{n}\bu|^{2}dV\,,
\ee
from which we can write \cite{DG1995}
\bel{fgt1a}
\shalf \dot{H}_{n} \leq -\nu H_{n+1} + c_{n} \|\nabla\bu\|_{\infty}H_{n}\,.
\ee
For simplicity, we have omitted the forcing. An integer-$D$-dimensional Gagliardo-Nirenberg inequality gives
\bel{fgt1b}
\|\nabla\bu\|_{\infty}\leq c_{n}H_{n+1}^{a/2}H_{1}^{(1-a)/2}
\ee
with $a=D/2n$ \& $n > D/2$. After re-arrangement, (\ref{fgt1a}) becomes 
\beq{fgt2}
\shalf \dot{H}_{n} &\leq& -\nu\left(1 - \shalf a\right)H_{n+1} + c_{n}\nu^{-\frac{a}{1-a}}H_{n}^{\frac{2}{2-a}}H_{1}^{\frac{1-a}{2-a}}\nonumber\\
&\leq& -\shalf \nu\left(1 - \shalf a\right)H_{n+1} + c_{n}\nu^{-\frac{a}{1-a}}H_{n}^{\frac{4n}{4n-D}}H_{1}^{\frac{2n-D}{4n-D}}\,.
\eeq
Divide by $H_{n}^{n\alpha_{n,1}}$ \& time average to give
\beq{fgt3}
\left<\frac{H_{n+1}}{H_{n}^{n\alpha_{n,1}}}\right>_{T} &\leq& c_{n}\nu^{-\frac{1}{1-a}}
\left<H_{n}^{\frac{n(4-D)\alpha_{n,1,D}}{4n-D}}H_{1}^{\frac{2n-D}{4n-D}}\right>_{T}\nonumber\\
&\leq& c_{n}\nu^{-\frac{1}{1-a}}\left<H_{n}^{\shalf (4-D)\alpha_{n,1,D}}\right>_{T}^{\frac{2n}{4n-D}}
\left<H_{1}\right>_{T}^{\frac{2n-D}{4n-D}}\,.
\eeq
Then a Holder inequality gives 
\beq{fgt4}
\left<H_{n+1}^{\shalf(4-D)\alpha_{n+1,1,D}}\right> &\leq &
\left<\frac{H_{n+1}}{H_{n}^{n\alpha_{n,1}}}\right>^{\shalf(4-D)\alpha_{n+1,1,D}}\\
&\times&
\left<H_{n}^{\frac{\shalf(4-D)n\alpha_{n,1,D}\alpha_{n+1,1,D}}{1-\shalf(4-D)\alpha_{n+1,1,D}}}\right>_{T}^{1-\shalf(4-D)\alpha_{n+1,1,D}}\,.\nonumber
\eeq
It is then easy to show that the exponent of $H_{n}$ within the average can be simplified to
\bel{fgt5}
\frac{\shalf(4-D)n\alpha_{n,1,D}\alpha_{n+1,1,D}}{1-\shalf(4-D)\alpha_{n+1,1,D}} = \shalf (4-D)\alpha_{n,1,D}\,.
\ee
Taking (\ref{fgt4}) \& (\ref{fgt5}) together \& using the dimensionless notation of $F_{n,m,D}$, we end up with
\bel{fgt6}
\left<F_{n+1,1}^{(4-D)\alpha_{n+1,1,D}}\right>_{T} \leq  c_{n,1}\left<F_{n,1}^{(4-D)\alpha_{n,1,D}}\right>_{T} 
+ c_{n,2}\left<F_{1,1,D}^{2}\right>_{T}\,.
\ee
To begin an iteration procedure it is necessary to have a bound in the $n=2$ case because $n > D/2$ \& $D = 2,\,3$. We  repeat the argument above for $n=2$ only
\bel{fgt7a}
\shalf\dot{H}_{1} \leq -\nu H_{2} + \|\bom\|_{4}^{2}\|\bom\|_{2}
\ee
We note that in $D$-dimensions $\|\bom\|_{4} \leq c\,\|\nabla\bom\|_{2}^{a}\|\bom\|_{2}^{1-a}$ were $a= D/4$. 
Thus we have
\bel{fgt7b}
\shalf\dot{H}_{1} \leq -\nu \left(1 - \quart D\right)H_{2} + c\,\nu^{-\frac{D}{4-D}}H_{1}^{\frac{6-D}{4-D}}
\ee
Firstly we consider
\beq{fgt7c}
\left<H_{2}^{\shalf\alpha_{2,1,D}}\right>_{T} &=& 
\left<\left(\frac{H_{2}}{H_{1}^{\beta}}\right)^{\shalf\alpha_{2,1,D}}H_{1}^{\shalf\beta\alpha_{2,1,D}}\right>_{T}\nonumber\\
&\leq&
\left<\frac{H_{2}}{H_{1}^{\beta}}\right>_{T}^{\shalf\alpha_{2,1,D}}
\left<H_{1}^{\frac{\shalf\beta\alpha_{2,1,D}}{1-\shalf\alpha_{2,1,D}}}\right>_{T}^{1-\shalf\alpha_{2,1,D}}
\eeq
Thus we must choose $\beta$ to make the exponent of $H_{1}$ equal to unity\,:
\bel{fgt7d}
\beta = 2\alpha_{2,1,D}^{-1} - 1 = \frac{6-D}{4-D}-1 = \frac{2}{4-D}
\ee
To see about the ratio we look at (\ref{fgt7a}) \& divide by $H_{1}^{\beta}$ to obtain
\bel{fgt8a}
\left<\frac{H_{2}}{H_{1}^{\beta}}\right>_{T} \leq c\,\nu^{-\frac{4}{4-D}}\left<H_{1}^{\frac{6-D}{4-D} - \beta}\right>_{T} = \nu^{-\frac{4}{4-D}}\left<H_{1}\right>_{T}\,.
\ee
Thus $\left<F_{2,1}^{(4-D)\alpha_{2,1,D}}\right>_{T} < \infty$. Then, from (\ref{fgt6}), the result follows for all $n \geq 1$
\bel{fgt9}
\left<F_{n,1,D}^{(4-D)\alpha_{n,1,D}}\right>_{T} \leq c_{n,1}Re^{3}\,.
\ee
Formally this is the equivalent of the result in \cite{FGT1981} when $D=3$.

\section{\small The result for $1 \leq m \leq \infty$}\label{gnisub}

The bound in (\ref{fgt9}) is true for $m=1$ only. To move up to the $m > 1$ case we use a Gagliardo-Nirenberg inequality in integer-$D$-dimensions
\bel{gni1}
\|A\|_{2m} \leq c\, \|\nabla^{N}A\|_{2}^{a}\|A\|_{2}^{1-a} 
\ee
where $2aN = D(m-1)/m$. Therefore, with $A \equiv \nabla^{n}\bu$, we use the $F_{n,m,D}$-notation. We also keep in mind the result (\ref{fgt9}) above to find
\beq{gni2}
\left<F_{n,m,D}^{(4-D)\alpha_{n,m,D}}\right>_{T} &\leq& 
c\,\left<F_{N+n,1,D}^{a(4-D)\alpha_{n,m,D}}F_{n,1,D}^{(4-D)\alpha_{n,m,D}(1-a)}\right>_{T}\nonumber\\
&=& 
c\,\left<\left(F_{N+n,1,D}^{(4-D)\alpha_{N+n,1,D}}\right)^{\frac{a\alpha_{n,m,D}}{\alpha_{N+n,1,D}}}
F_{n,1,D}^{(4-D)(1-a)\alpha_{n,m,D}}\right>_{T}\\
&\leq&\left<F_{N+n,1}^{(4-D)\alpha_{N+n,1,D}}\right>_{T}^{\frac{a\alpha_{n,m,D}}{\alpha_{N+n,1,D}}}
\left<F_{n,1,D}^{\frac{(4-D)\alpha_{n,m,D}(1-a)\alpha_{N+n,1,D}}
{\alpha_{N+n,1,D} - a\alpha_{n,m,D}}}\right>_{T}^{1 - \frac{a\alpha_{n,m,D}}{\alpha_{N+n,1,D}}}\,.\nonumber
\eeq
Using the fact that $2aN = D(m-1)/m$ \& the expression for $\alpha_{n,m,D}$ given in (\ref{alphad1}), we can then show that the exponent of $F_{n,1,D}$ in the time average satisfies
\bel{gni3}
\frac{(4-D)\alpha_{n,m,D}(1-a)\alpha_{N+n,1,D}}
{\alpha_{N+n,1,D} - a\alpha_{n,m,D}} = (4-D)\alpha_{n,1,D}\,.
\ee
Using (\ref{fgt9}), we see that both factors on the right hand side of (\ref{gni2}) are bounded \& give (\ref{FnmDbnd}). \hfill $\blacksquare$


\end{document}